\documentclass[twocolumn,prl,showpacs,amsmath,amssymb]{revtex4}
\usepackage{graphicx}
\usepackage{dcolumn}
\usepackage{bm}
\usepackage{graphicx,color}
\definecolor{r}{rgb}{1,0,0}
\definecolor{g}{rgb}{0,1,0}
\definecolor{b}{rgb}{0,0,1}

\begin{document}

\title{Drag force scaling for penetration into granular media}

\author{Hiroaki Katsuragi$^{1,2}$ and Douglas J. Durian$^1$}
\affiliation{$^1$Department of Physics and Astronomy, University of Pennsylvania, Philadelphia, PA 19104, USA \\
$^2$Department of Earth and Environmental Sciences, Nagoya University,  Nagoya 464-8601, Japan}
\date{\today}

\begin{abstract}
Impact dynamics is measured for spherical and cylindrical projectiles of many different densities dropped onto a variety non-cohesive  granular media.  The results are analyzed in terms of the material-dependent scaling of the inertial and frictional drag contributions to the total stopping force.   The inertial drag force scales similar to that in fluids, except that it depends on the internal friction coefficient.  The frictional drag force scales as the square-root of the density of granular medium and projectile, and hence cannot be explained by the combination of granular hydrostatic pressure and Coulomb friction law.  The combined results provide an explanation for the previously-observed penetration depth scaling.
\end{abstract}

\pacs{45.70.Ht, 45.70.Cc, 83.80.Fg, 89.75.Da}

\maketitle

Collision is one of the most fundamental processes in nature, and can be exploited to uncover the basic physics of systems ranging from planets to elementary particles.  Impact of projectiles into a pack of grains has been of increasing interest for highlighting and elucidating the unusual mechanical properties of granular materials. This includes studies of crater morphology~\cite{Amato1998a, Walsh2003a, Zheng2004a}, penetration depth~\cite{Uehara2003a, Ambroso2005a, deBruyn2004a}, dynamics~\cite{Daniels2004a, Ciamarra2004a, Lohse2004a, Ambroso2005b, Hou2005a, Tsimring2005a, Katsuragi2007a, Goldman08, Goldman10, Vazquez2011, BehringerPRL12, Behringer13}, boundary effects~\cite{BoudetPRL06, Nelson2008, Seguin2008, vonKann2010,  Vazquez2011, AlshulerRCF12} and packing-fraction effects~\cite{Goldman10, Royer2011, AlshulerRCF12}.   One finding is that the penetration depth scales as $d\sim {D_p}^{2/3}H^{1/3}$ where $D_p$ is projectile diameter and $H$ is the total drop distance~\cite{Uehara2003a, Ambroso2005a}.  Granular impact is also important for military and industrial applications \cite{roddy, zukas}, and cone penetration tests are used for {\it in-situ} soil characterization~\cite{Yu1998a, LosertGM07}.

In a previous study~\cite{Katsuragi2007a} we measured the impact dynamics of a 2.54~cm steel ball onto a packing of glass beads.  The stopping force was found to be the sum of an inertial drag, proportional to the square of the speed $v$, and a frictional drag, proportional to depth $z$.  This has since been supported by several other experiments~\cite{Seguin2008, Goldman08, BrzinskiSM10, vonKann2010, Vazquez2011, AlshulerRCF12, BehringerPRL12, Behringer13}.  The equation of motion during impact is thus
\begin{equation}
ma = -mg + mv^2/d_1 + k|z|,
\label{eq:force_model}
\end{equation}
where $m$ is projectile mass, $g=980$ cm/s$^2$, and $d_1$ and $k$ are materials parameters expressed with units of a length and a spring constant, respectively.  By using $ma={\rm d}K/{\rm d}z$ and an integrating factor, as in Ref.~\cite{Ambroso2005b}, this can be solved for speed versus depth:
\begin{equation}
\frac{v^2}{v_0^2} = e^{-\frac{2z}{d_1}}-\frac{k d_1 z}{m v_0^2}
   +\left( \frac{g d_1}{v_0^2}+\frac{k d_1^2}{2 m v_0^2}\right) \left(1- e^{-\frac{2z}{d_1}} \right),
\label{eq:force_model_solution}
\end{equation}
where $v_0$ is the impact speed at $z=0$.  The final penetration depth $d$ is then given by the limit of $v \to 0$ as 
\begin{equation}
	\frac{2d}{d_1} = 1  +   \frac{2mg}{kd_1}
	+ W\left( \frac{2 m v_0^2 - 2mgd_1-kd_1^2}{ kd_1^2 \ e^{1+2mg/ k d_1}   } \right),
\label{eq:d_solution}
\end{equation}
where $W(x)$ is the Lambert $W$-function.  An additional constant stopping force $f_0$ \cite{deBruyn2004a, Behringer13} can be included in these expressions by replacing $g$ with $g-f_0/m$.

Some open questions are how $d_1$ and $k$ scale with the materials properties of the projectile and the granular packing, and how this conspires to give $d\sim D_p^{2/3}H^{1/3}$.  For inertial drag, exactly as for hydrodynamic drag at high Reynolds number Re, momentum transfer gives the expectation $mv^2/d_1 \sim A \rho_g v^2$ where $A$ is the projected projectile area and $\rho_g$ is the mass density of the granular medium.   For frictional drag, the combination of hydrostatic pressure and Coulomb friction gives $k|z| \sim \mu g \rho_g A |z|$ where the internal frictional coefficient is $\mu = \tan \theta_r$ and $\theta_r$ is the angle of repose.  The scaling of $d_1$ and $k$ would thus be
\begin{eqnarray}
	d_1 / D_p &\sim& \rho_p/\rho_g, \label{eq:d1expectation} \\
	kD_p / mg &\sim& \mu \rho_g / \rho_p, \label{eq:kexpectation}
\end{eqnarray}
where $D_p$ and $\rho_p$ correspond to diameter and density of projectile, respectively. Here, we test the speed versus depth prediction of Eq.~(\ref{eq:force_model_solution}), and we compare fitted values of $d_1$ and $k$ with the above expectations.  While neither turns out to be quite correct, we connect the results to the observed penetration depth scaling.

Our basic experimental setup is identical with previous work~\cite{Katsuragi2007a}.   The velocity of the projectile $v(t)$ at time $t$ is computed by particle-image velocimetry applied to fine stripes on a vertical rod glued to the top of the projectile.  The system has 20~$\mu$s temporal resolution and 100~nm  spatial resolution, which is fine enough to compute position and acceleration from $v(t)$.   The primary difference from Ref.~\cite{Katsuragi2007a} is that now we vary both the projectiles and the granular medium.  We begin with 0.35~mm diameter glass beads, prepared to a reproducible random packing state by slowly turning off a fluidizing up-flow of air.  Into this we drop a wide variety of projectiles, listed in Table~\ref{tab:projectile}, from free-fall heights between 0 to 85~cm.  The first type is spheres.  The second type is aluminum cylinders, which are dropped lengthwise with the axis horizontal and parallel to the surface of the granular medium.  In both cases the effective density is given by projectile plus rod mass divided by projectile volume.  The length of the cylinders is varied, but we find that it does not affect the dynamics or final penetration depths.  

\begin{table}
\begin{ruledtabular}
\begin{tabular}{lccc}
Projectile & $\rho_p$ (g/cm$^3$) & $D_p$ (cm) & $L_p$ (cm) \\
\colrule
Hollow PP ball & 0.51 & 2.54 & - \\
Wood ball & 0.95 & 2.54 & - \\
Delrin ball & 1.65 & 2.54 & - \\
PTFE ball & 2.46 & 2.54 & - \\
Steel ball & 7.96 - 159 & 0.3175 - 5.08 &- \\
Tungsten carbide ball & 15.3 & 2.54 & - \\
Aluminum cylinder & 2.89 - 4.26 & 0.635 -1.27 & 5.08 - 15.24 \\
\end{tabular}
\end{ruledtabular}
\caption{Projectile properties. The steel sphere diameters are $D_p=5.08$, $2.54$, $1.27$, $0.635$, and $0.3175$ cm.  The aluminum cylinders diameters are $D_p=1.27$ and $0.635$~cm; each have lengths $L_p=5.08$, $10.16$, and $15.24$~cm. The density $\rho_p$ is projectile plus rod mass  divided by projectile volume.}
\label{tab:projectile}
\end{table}

\begin{figure}
\includegraphics[width=3.00in]{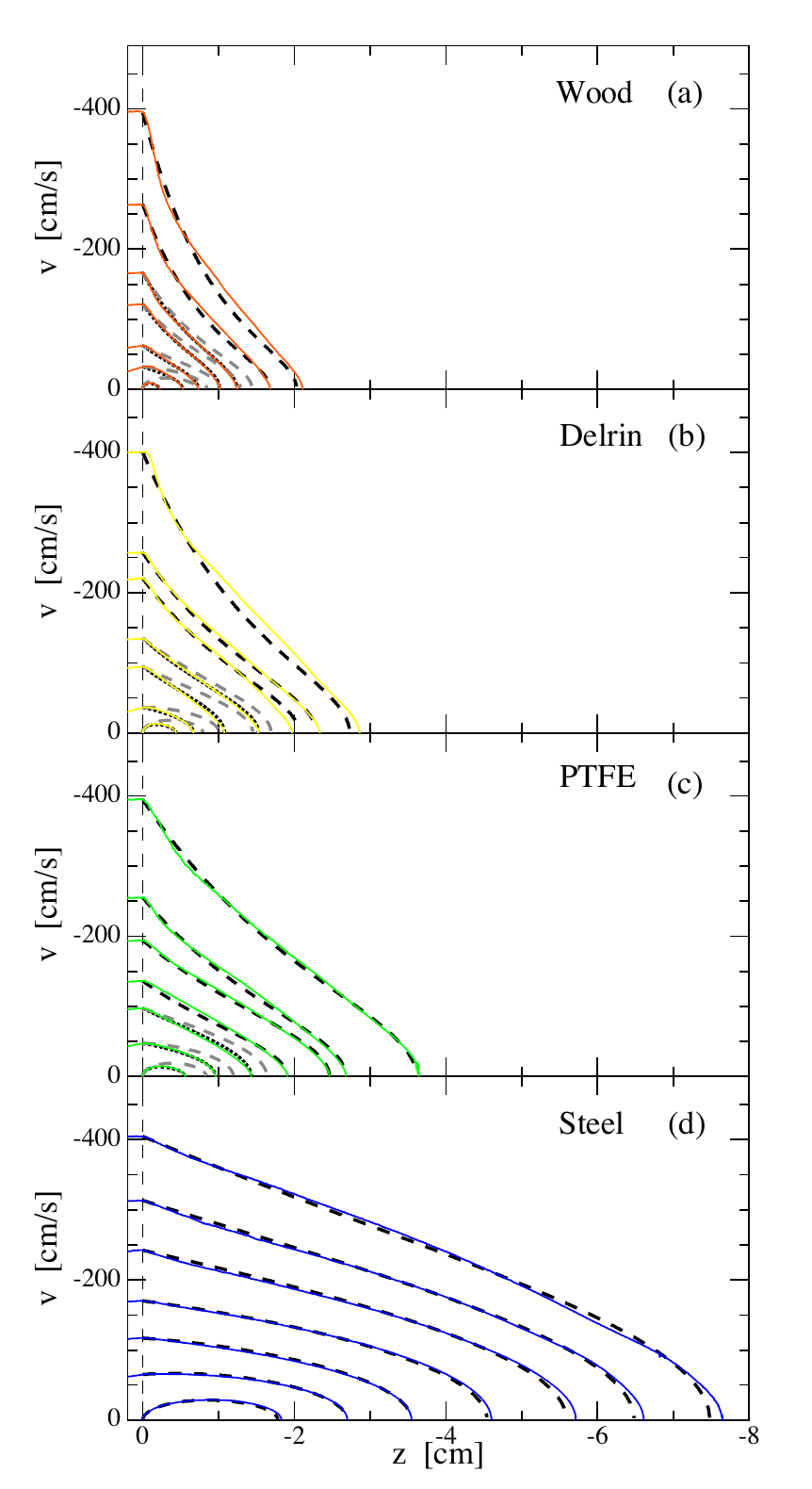}
\caption{(color online) Example velocity versus depth data for $D_p=2.54$~cm diameter spheres.  The impact begins at $z=0$, and proceeds downward in the $-z$ direction.   The black dashed curves are a simultaneous fit to Eq.~(\ref{eq:force_model_solution}), where a single pair of $k$ and $d_1$ values is found for different initial impact speeds.  The grey dotted curves are also fits to Eq.~(\ref{eq:force_model_solution}), but where $k$ and $d_1$ are adjusted for each impact speed and hence do not represent well-defined materials parameters.}
\label{fig:v_z}
\end{figure}

Example speed versus position data are plotted in Fig.~\ref{fig:v_z} for $D_p=2.54$~cm diameter projectiles of four different densities, dropped onto the glass bead packing, with initial impact speeds ranging from zero to 400~cm/s.   For slow impacts, the speed first increases and then decreases with depth.  For faster impacts, the speed versus depth curves gradually change from concave down to concave up.  Generally, there is a rapid decrease of speed to zero at the final penetration depth. We obtain good fits to these complex trajectories by adjusting $k$ and $d_1$, as shown in Fig.~\ref{fig:v_z}.  The displayed level of agreement is typical for all projectiles, including the cylinders.

\begin{figure}
\includegraphics[width=3.00in]{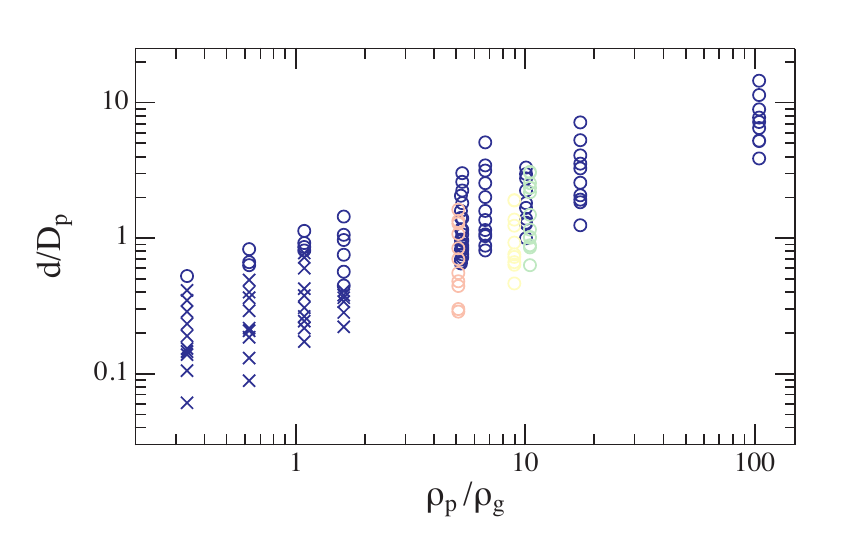}
\caption{(color online) Scatter plot of penetration depth versus projectile density, scaled respectively by projectile diameter and grain density. An open circle is plotted for conditions such that the $k$ and $d_1$ fitting parameters are constant; a cross is plotted otherwise.   Blue is for glass beads; green is for rice; pink is for beach sand; yellow is for sugar.}
\label{PD}
\end{figure}

Unfortunately a good simultaneous fit for a given projectile to a single pair of $k$ and $d_1$ values at all impact speeds can be obtained only for denser projectiles.  For the less dense projectiles, the fitting parameters become constant only at high impact speeds.  Then the same values for $k$ and $d_1$ are obtained as from the analysis method of Ref.~\cite{Katsuragi2007a}.  This holds roughly for $d>D_p/2$ and $\rho_p>2\rho_g$ as shown in Fig.~\ref{PD}.  We speculate that for low density projectiles and small impact speeds, the penetration can be shallow enough that the detailed shape of projectile must be taken into account \cite{Newhall2003a}.  In this regime surface flows and surface roughness could plays a role, though identical penetration behavior was found earlier for slick and tacky projectiles of the same size and density \cite{Uehara2003a}.  We note that including a constant force $f_0$ as a third free parameter does not noticeably change the fits; the largest fit value is $f_0/m=2.4$~cm/s$^2$, which is small compared to $g$.  We note, too, that $f_0$ cannot be chosen such that $k$ and $d_1$ become constant.  For the rest of the paper, we restrict attention to conditions where the deduced $k$ and $d_1$ values do not depend on impact speed and hence can be considered as materials parameters.

\begin{figure}
\includegraphics[width=3.00in]{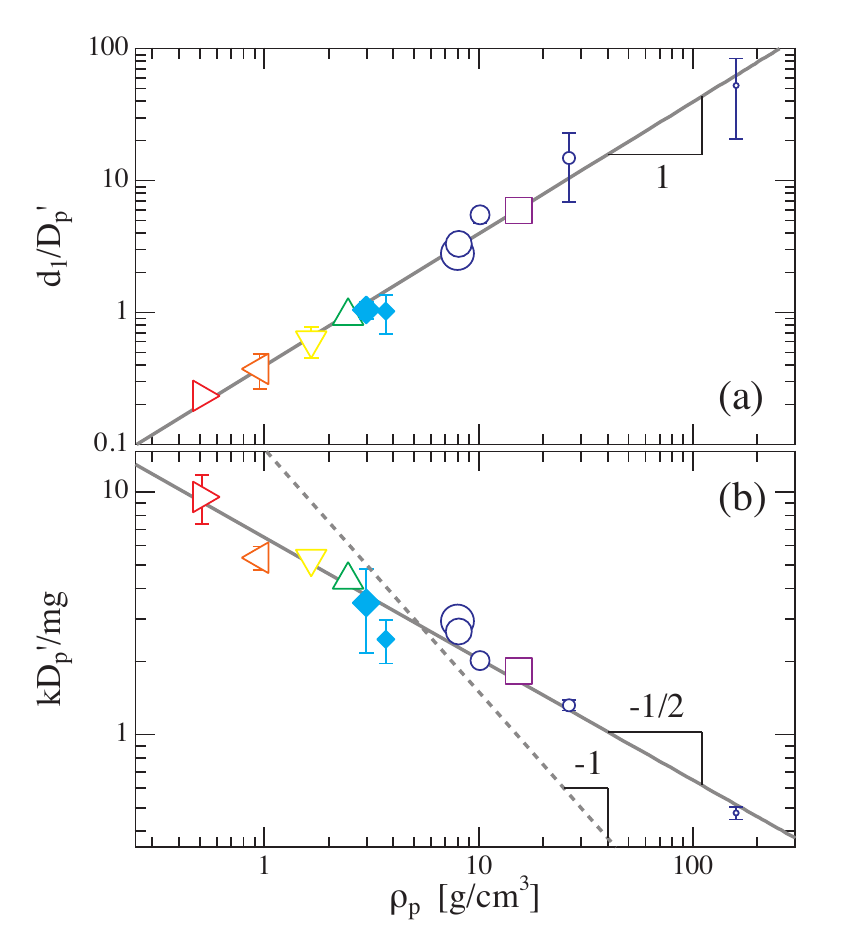}
\caption{(color online)  (a) Dimensionless inertial drag length scale $d_1/D_p'$, and (b) dimensionless quastistatic drag coefficient $kD_p'/mg$, versus projectile density, for impact into glass beads.  Here $D_p'$ is the effective projectile diameter; $m$ is its total mass, including rod; $\rho_p$ is $m$ divided by projectile volume; $g=980$~cm/s$^2$.   Open symbols are for spheres made of hollow PP ($\rhd$), wood ($\lhd$), PTFE ($\bigtriangledown$), delrin ($\bigtriangleup$), steel ($\bigcirc$), and tungsten carbide ($\square$);  closed diamonds are for aluminum cylinders.  Symbol sizes increase with $D_p'$, with values given in Table~\ref{tab:projectile}.  The solid gray lines denote power-laws as labeled.  The dashed line in (b) is the expected scaling, $k D_p'/mg \sim \rho_p^{-1}$.}
\label{fig:rho_scale}
\end{figure}

We now compare the fitting parameters with Eqs~(\ref{eq:d1expectation}-\ref{eq:kexpectation}).  The expectation for $d_1$ is based on momentum transfer, so that the inertial drag force is $mv^2/d_1 \sim A \rho_g v^2$ just like an object moving in a fluid at high Reynolds number.  For spheres and cylinders, the characteristic length is thus $d_1\sim m/A\rho_g$ and can be written as $d_1 \sim D_p' \rho_p/\rho_g$ if we take $D_p'$ to be 1 times diameter for spheres and $3\pi/8$ times diameter for cylinders (the cylinder length drops out of the ratio $m/A$).   For a unified analysis, we therefore divide the fitted value of $d_1$ by $D_p'$ and plot versus $\rho_p$ in Fig.~\ref{fig:rho_scale}a.  We find that all data, including the cylinder data, collapse nicely onto a power-law with the expected slope of one; i.e. $d/D_p' \sim \rho_p^1$ holds as per expectation.

For the fitting parameter $k$ that sets the quasi-static friction force, $k|z|$, we now make a similar comparison with expectation by plotting $kD_p'/mg$ versus $\rho_p$ in Fig.~\ref{fig:rho_scale}b.  All the data, including the cylinder, again collapse nicely to a power law in projectile density.  However, the expected power-law $kD_p/mg\sim1/\rho_p$ is clearly wrong.  The data are instead consistent with $kD_p/mg\sim1/\sqrt{\rho_p}$.  Therefore, the nature of the quasistatic frictional drag is different from the simple combination of Coulomb friction and hydrostatic pressure.

\begin{figure}
\includegraphics[width=3.00in]{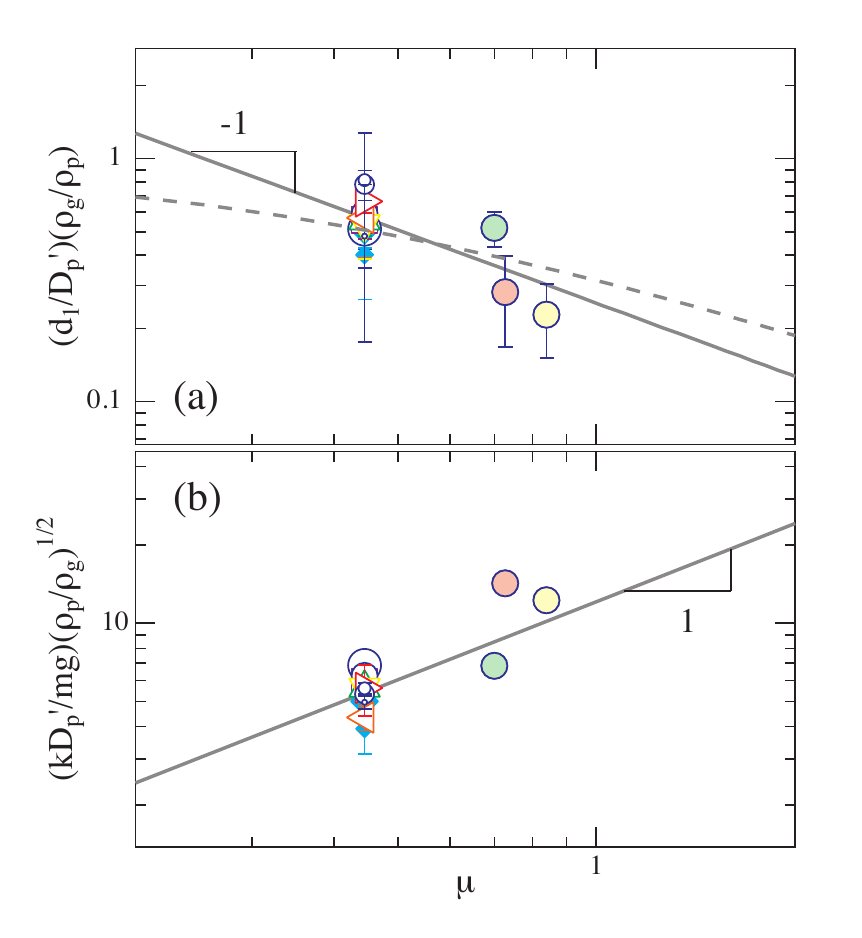}
\caption{(color online).  (a) $(d_1/D_p')(\rho_g/\rho_p)$ and (b) $(kD_p'/mg)(\rho_p/\rho_g)^{1/2}$, versus $\mu=\tan\theta_r$, where $\theta_r$ is the repose angle of the granular medium. The symbols at $\mu=0.45$ are for glass beads, with the same symbol codes as in Fig.~\ref{fig:rho_scale}.  The colored circles at increasing $\mu$ are for rice (green), beach sand (pink), and sugar (yellow).  The gray lines indicate power-law behavior.  The dashed curve in (a) is a fit to $d_1 \propto 1/(1+\alpha\mu)$, giving $\alpha=2.2\pm0.6$.}
\label{fig:mu_scale}
\end{figure}

To investigate this further, we perform a second series of experiments where $D_p$=2.54~cm diameter steel spheres are dropped into rice, beach sand, and sugar (with materials properties listed in Table~\ref{tab:granu}).  As in Fig.~\ref{fig:v_z}, speed vs position data for a range of impact speeds are fit to Eq.~(\ref{eq:force_model_solution}) to obtain values of $d_1$ and $k$.  Based on Fig.~\ref{fig:rho_scale}, and assuming that the x-axis of this figure is correctly made dimensionless by dividing out the bulk density $\rho_g$ of the granular medium, the observed scalings so far are $d_1/D_p' \sim (\rho_p/\rho_g)$ and $kD_p'/mg \sim (\rho_p/\rho_g)^{-1/2}$.  Therefore, we divide out this density dependence and plot it in Fig.~\ref{fig:mu_scale} versus the only remaining material property of the grains: the internal friction coefficient $\mu=\tan\theta_r$ where $\theta_r$ is the draining angle of repose.  In this figure the data for glass beads, from Fig.~\ref{fig:rho_scale}, all lie at $\mu=0.45$.  The range of $\mu$ values is less than a factor of two, but to within uncertainty the scaled $d_1$ and $k$ parameters collapse to power-laws in $\mu$.  For the quasistatic frictional drag coefficient we find $k\sim\mu$, which is expected for Coulomb friction.  For the speed-squared inertial drag coefficient, we find $d_1 \sim 1/\mu$.  Therefore the inertial drag force is proportional to $\mu$, which could correspond to an added-mass effect whereby the volume of grains boosted to the projectile speed grows in proportion to $\mu$.  This is unlike the case of simple fluids, where the fluid flow and the inertial drag force at high Re depend only on the density of the fluid.

As an alternative analysis for the $\mu$ dependence of $d_1$, one could imagine that an inertial drag force $\sim \rho_g v^2 A$ loads the contacts and gives an additional friction force of $\mu$ times this loading.  Then the total velocity-squared force is $m v^2/d_1 \sim (1+\alpha\mu)\rho_g v^2 A$, which gives $d_1\propto 1/(1+\alpha\mu)$.  This reasonably fits the data, as shown in Fig.~\ref{fig:mu_scale}a with $\alpha=2.2\pm0.6$.  The residuals are smaller for the power-law form.

\begin{table}
\begin{ruledtabular}
\begin{tabular}{lccc}
Granular Material & $\rho_g$ (g/cm$^3$) & $\theta_r$  & Grain size (mm)\\
\colrule
Glass beads & 1.45 & $24^\circ$ & 0.25 - 0.35 \\
Rice & 0.77 & $35^\circ$ & 2$\times$8 \\
Beach sand &1.51 & $36^\circ$ & 0.2 - 0.8 \\
Sugar & 0.89 & $40^\circ$ & 0.4 - 0.7 \\
\end{tabular}
\end{ruledtabular}
\caption{Granular media properties: $\rho_g$ is bulk density; $\theta_r$ is drainage angle of repose; size is the range of grain diameters, except for rice where it is the length of short and long axes.}
\label{tab:granu}
\end{table}

Combining the power-laws in Figs.~\ref{fig:rho_scale}-\ref{fig:mu_scale}, and the actual numerical prefactors, the inertial and frictional drag coefficients are altogether found by measurements for a range of projectiles and grains to be consistent with
\begin{eqnarray}
d_1/ D_p' &=& (0.25/\mu)(\rho_p/\rho_g) \label{eq:d1_scale}, \\
kD_p' / mg & =& 12 \mu \left( \rho_g / \rho_p \right)^{1/2}. \label{eq:km_scale}
\end{eqnarray}
The two main differences from the simple expectation of Eqs.~(\ref{eq:d1expectation}-\ref{eq:kexpectation}), are the factor of $1/\mu$ in $d_1$ and the density ratio exponent of 1/2 rather than 1 in $k$.  These results may be inspected differently by re-writing the equation of motion as
\begin{equation}
ma = -mg + 2.7 \mu \rho_g v^2A + 8.0 \mu  \left( \rho_p \rho_g \right)^{1/2}g|z|A.
\label{eq:universal_force_model}
\end{equation} 
Note that the numerical prefactor for the depth-dependent frictional drag is significantly larger than unity, and that $\sqrt{\rho_p\rho_g}$ is larger than $\rho_g$ for dense projectiles.  For both reasons,  frictional drag exceeds the value expected from hydrostatic pressure and Coulomb friction.  One might have expected the opposite effect, either by a decrease in contact area between projectile and grains due to ejection of grains or by fluidization of the grains from the motion of the projectile.  Our results instead appear to indicate that frictional contacts are loaded by the motion of the projectile, so that the medium is stronger than set by gravity alone.  Such behavior is not seen for the fast horizontal rotation of bars \cite{BrzinskiSM10}, where the depth- and speed-dependent forces are easily disentangled, and warrants further attention.

\begin{figure}
\includegraphics[width=3.00in]{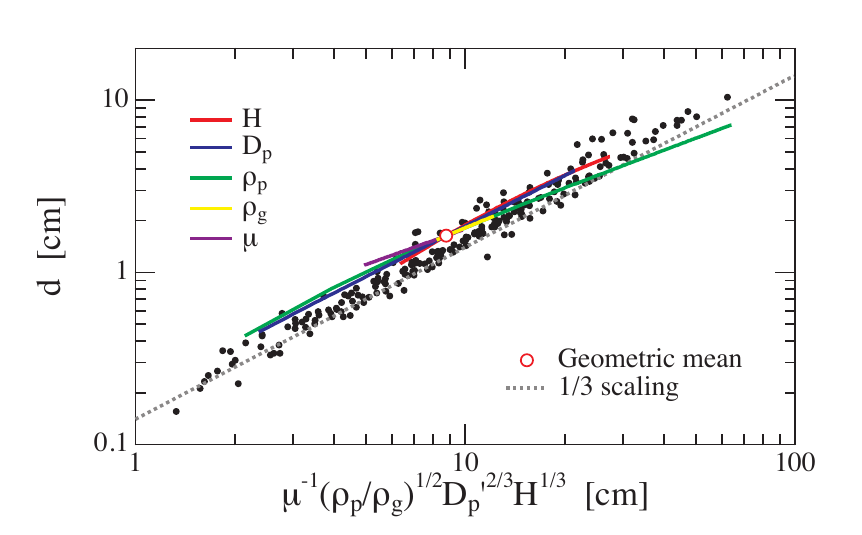}
\caption{(color online). Final penetration depth scaling plot. All trials data are shown and they reasobaly agree with the empirical scaling $d=0.14\mu^{-1}(\rho_p/\rho_g)^{1/2}D_p'^{2/3}H^{1/3}$ (dotted line)~\cite{Uehara2003a,Ambroso2005a}. Colored curves show the force law predictions. They are also close to data and the empirical scaling.}
\label{fig:H_scale}
\end{figure}

As a final test we now compare data for the final penetration depth $d$ with Eq.~(\ref{eq:universal_force_model}).  Prior observations \cite{Uehara2003a,Ambroso2005a} are consistent with the empirical form
\begin{equation}
d=0.14\mu^{-1}(\rho_p/\rho_g)^{1/2}D_p'^{2/3}H^{1/3}.
\label{eq:d_scaling}
\end{equation} 
Thus in Fig.~\ref{fig:H_scale} we plot penetration depth data for all trials versus the quantity $(\rho_p/\rho_g)^{1/2}D_p'^{2/3}H^{1/3}$ from the right-hand side of this expression.  This collapses our new data to within the experimental uncertainty, including that for the cylinder, to the line $y=0.14x$ and hence shows agreement with prior work.  However Eqs.~(\ref{eq:d_solution},\ref{eq:universal_force_model}) do not predict perfect power-law behavior.  To compare with data, we first calculate the geometric mean of each of the five variables $H$, $D_p$, $\rho_p$, $\rho_g$, and $\mu$ over the range of experimental conditions employed here.  The predicted penetration depth for these mean values is shown by a single red open circle in Fig.~\ref{fig:H_scale}.  Then we vary each of variables, one at a time, keeping all others fixed at their mean value.  The resulting five predicted penetration depth curves are included in Fig.~\ref{fig:H_scale}.  They are not identical, or even perfect power laws, but are all close together and in fair agreement with the data.  Thus the empirical penetration depth data scaling is satisfactorily understood in terms of the nature of the stopping force exerted by the medium onto the projectile.

In conclusion, we developed an exact solution of Eq.~(\ref{eq:force_model}) for the dynamics of penetration, and we conducted a wide range of granular impact experiments to elucidate the materials dependence of the inertial and frictional contributions to the stopping force.  The final equation of motion, Eq.~(\ref{eq:universal_force_model}), is roughly consistent with the empirical scaling of penetration depth versus drop height and materials parameters.  However it is not consistent with the apparently simplistic expectation of Eqs.~(\ref{eq:d1expectation}-\ref{eq:kexpectation}).  So there is unanticipated physics, yet to be understood, which could possibly arise from motion-loading of frictional contacts or from granular flow fields that depend on the internal friction coefficient.

This work was supported by the JSPS Postdoctoral Fellowships for Research Abroad (HK) and the National Science Foundation through Grant Nos. DMR-0704147 and DMR-1305199.

\bibliography{impact2}

\end{document}